\documentclass[intlimits,twoside,a4paper]{article}

\usepackage[utf8]{inputenc}

\usepackage{mathrsfs}
\usepackage{bbold}
\usepackage{float}
\usepackage{bm}
\usepackage{amsmath} 
\usepackage{epsfig, bm, amsfonts} 
\usepackage{hhline}
\usepackage[textsize=tiny]{todonotes}
\usepackage[greek,english]{babel}
\usepackage{teubner}
\usepackage{graphicx}
\RequirePackage{color}

\usepackage[eqsecnum]{cmpj3}

\newcommand{\be}{\begin{equation}}
\newcommand{\ee}{\end{equation}}
\newcommand{\bea}{\begin{eqnarray}}
\newcommand{\eea}{\end{eqnarray}}
\newcommand{\bem}{\begin{matrix}}
\newcommand{\eem}{\end{matrix}}
\newcommand{\nnb}{\nonumber}

\def\koppa{\hbox{\foreignlanguage{greek}{\coppa}}}
\def\smallkoppa{{\hbox{\foreignlanguage{greek}{\footnotesize\coppa}}}}

\issue{2023}{26}{3}{33606}
\doinumber{10.5488/CMP.26.33606}

\title{The fifty-year quest for universality in percolation theory in high dimensions}
\author[T. Ellis, R. Kenna, B. Berche]{T. Ellis\orcid{0000-0003-4713-7875}\refaddr{label1}, 
	R. Kenna\orcid{0000-0001-9990-4277}\refaddr{label2},
 	B. Berche\orcid{0000-0002-4254-807X}\refaddr{label3} 
 }
\addresses{
\addr{label1} Applied Mathematics Research Centre, Coventry University, England
\addr{label2} Applied Mathematics Research Centre, Coventry University, England
	and L4 collaboration, Leipzig-Lorraine-Lviv-Coventry, Europe
\addr{label3} Laboratoire de Physique et Chimie Th\'eoriques,  CNRS - Universit\'e de Lorraine, UMR 7019,
    	 Nancy, France 
	 	and L4 collaboration, Leipzig-Lorraine-Lviv-Coventry, Europe
}

\date{Received May 31, 2023, in final form June 21, 2023}

\begin{document}
\maketitle

\begin{abstract}
Although well described by mean-field theory in the thermodynamic limit,  scaling  has long been puzzling for finite systems in high dimensions.
This raised questions about the efficacy of the renormalization group and foundational concepts such as universality, finite-size scaling and hyperscaling, until recently believed not to be applicable above the upper critical dimension.
Significant theoretical progress has been made resolving these issues, and  tested in numerous simulational studies of spin models. 
This progress rests upon superlinearity of correlation length, a notion that for a long time encountered resistance but is now broadly accepted.
Percolation theory brings added complications such as proliferation of interpenetrating clusters in apparent conflict with suggestions coming from random-graph asymptotics and a dearth of reliable simulational guidance.
Here we report on  recent theoretical progress in percolation theory in the renormalization group framework in high dimensions that accommodates  superlinear correlation and renders most of the above concepts mutually com\-pa\-tible under different boundary con\-di\-tions. Results from numerical simulations for free and periodic boundary conditions which differentiate between previously competing theories are also presented. 
Although still fragmentary, these Monte Carlo results support the new framework which restores the renormalization group and foundational concepts on which it rests.  

 \end{abstract}

\section{Scaling in percolation theory}
\label{sec_1}

Although it can be traced back to Paul Flory's studies of gelation in the 1940's, the subject of percolation was introduced mathematically by Simon Broadbent and John Hammersley in 1957  \cite{BH}.
While these pursuits were motivated by interest in designing filters for gas masks and coal miners, the advent of renormalization group theory opened ways to  understand universality and its origins at a very fundamental (model independent) level.
Dietrich Stauffer's famous book \cite{Stauffer} describes a 1971 paper by John Essam and Keith Gwylyn \cite{EssG} as ``one of the starting points of the later avalanche of publications'' on the topic of percolation. 
Essam and Gwylyn had developed an analogy between percolation and ferromagnetism, drawn by Kees Fortuin and Piet Kasteleyn \cite{FK69} a few years previously. 
This paper spans this full period --- from the early days of renormalization group, through some of the challenges it faced to recent progress in resolving them.

Applications of percolation theory  extend far beyond gas masks and coal mining --- to chemistry, geology, network science, sociology, epidemiology and beyond. 
To quote Essam~\cite{Essam}
\begin{quote}
    ``A percolation model is a collection of points distributed in space, certain points of which are said to be linked. Whether or not two points are linked is governed by a random mechanism, the details of which depend on the context in which the model is used.''
\end{quote}
Here, we are interested in critical phenomena emerging from this random mechanism. 
The simplest approach is, of course, mean-field, and this is valid above the upper critical dimension where fluctuations play ``only'' irrelevant roles. 
The upper critical dimension in the case of percolation is $d_{\rm c}=6$.
Though the high-dimensional regime is not of direct importance for gas masks and coal mines, it is of great importance for the renormalization group (RG) because, although RG succeeded in describing critical phenomena below $d_{\rm c}$, it has been persistently problematic above $d_{\rm c}$ for finite-size systems
\cite{ourreview}.

Here we address finite-size scaling above the upper critical dimension for percolation theory.
Following Essam and Gwylyn, we draw on close parallels with ferromagnetism. 
To this end, we consider a lattice with sites or links deemed active with a given probability $p$. 
Proximity of these sites or bonds is deemed to build connected clusters which can carry current, liquid or any other signal, while percolation is the phenomenon which appears at a given probability, $p_\infty$, above which the signal crosses the whole system. 
The subscript means that we consider macroscopic properties, i.e., the thermodynamic limit, or the infinite volume limit.

In percolation theory, the  phase transition is second-order between two regimes (e.g., insulating vs conducting) and the order parameter, $P_\infty(p)$, which is the fraction of sites which belongs to a percolating cluster, has, close to the threshold, the  singular behavior
\be
P_\infty(p)\sim\left\{
\bem (p-p_\infty)^\beta, \hfill\ & \hbox{ if } p>p_\infty,\\
\ 0,\hfill\ & \hbox{ if } p<p_\infty.\eem
\right.
\label{eq-beta}
\ee
It is similar to the magnetization in a ferromagnet, with $p$ and $p_\infty$ analogous to associated temperature (or its inverse) and its  critical value, respectively. The critical exponent is  $\beta=1$ in mean-field theory (MFT).

The analog of the susceptibility is the average size of clusters (excluding infinite ones). 
As we shall remind the reader, it is also given by the second moment of the cluster number 
distribution
 and it behaves like
\be
 S_\infty(p)\sim |p-p_\infty|^{-\gamma},
 \label{eq-gamma}
\ee
with $\gamma=1$ in the mean-field regime. 
The related correlation function  is the probability that a site, which is a distance $r$ from an occupied site, belongs
to the same finite cluster. 
It behaves like $\exp(-r/\xi_\infty)$
with the correlation length diverging on approaching the percolation threshold as
\be
\xi_\infty(p)\sim|p-p_\infty|^{-\nu}
\label{eq-nu}
\ee
and with $\nu= 1/2$ in MFT.

The density of clusters of mass $s$ (i.e., the number of clusters of a given size), normalized per lattice site,  can be expressed as 
\be
n_\infty(s,p)\sim s^{-\tau}{\mathscr N}\left(\frac{s}{s^{\rm max}_\infty(p)}\right),
\label{eq-tau}
\ee
where $s^{\rm max}_\infty(p)$ is the typical mass of the largest cluster.
This diverges as the critical point is approached and we write 
\be
s^{\rm max}_\infty(p)\sim|p-p_\infty|^{- 1/\sigma}.\label{eq-sigma}
\ee
In MFT, the associated exponents are $\tau= 5/2$ and $\sigma = 1/2$. The  fractal dimension   of the large clusters is denoted by $D_{\rm f}$ and we write    $s^{\rm max}_\infty \sim\xi_\infty^{D_{\rm f}}$.
In the infinite volume limit, equations~(\ref{eq-nu}) and (\ref{eq-sigma}) imply 
\be
\sigma=\frac 1{\nu D_{\rm f}}.
\label{sigma}
\ee
There is, however, an analogue of the free energy density in percolation theory. 
This is the density of clusters
\be
K_\infty(p)=\sum_{s} n_\infty(s,p).
\label{eq-KSum}
\ee
As we shall soon see, this scales as the free energy density in field theory.
We write it as
\be
K_\infty(p)\simeq\int \rd s\ \!s^{-\tau}{\mathscr N}(s/s^{\rm max}(p))
=|p-p_\infty|^{({\tau-1})/\sigma}\int \rd\varsigma\ \!\varsigma^{-\tau}{\mathscr N}(\varsigma),
\label{Kinf}
\ee
where we set the variable $\varsigma = s/s_\infty^{\rm max}(p)$. 
Anticipating its free-energy role, we write the scaling form of the singular part of $K_\infty$ as {inverse} correlation volume
$
K_\infty(p)\sim \xi_\infty^{-d} \sim |p-p_\infty|^{d\nu} $.
This gives
\be
\tau=\frac d{D_{\rm f}}+1
\label{tau}
\ee
and allows us to define the critical exponent $\alpha$
\be
K_\infty(p)\sim  |p-p_\infty|^{2-\alpha},
\label{eq-alpha}
\ee
with $\alpha=-1$ in MFT.

The above critical exponents are interdependent.
Firstly, 
{if a site is occupied (probability $p$), 
it is either in an infinite percolating cluster (probability $P_\infty$) 
or in one of the other clusters [probability $ \sum_{s=1}^{s^{\rm max}}{s n_\infty (s,p)}$].}
For {associated probabilities} to be on the same scale,
equations~(\ref{eq-beta}), (\ref{eq-tau}) and (\ref{eq-sigma}) relate $\beta$, $\tau$ and $\sigma$.
Similarly, the mean size of finite clusters  
${\sum_{s=1}^{s^{\rm max}}{s^2 n_\infty (s,p)}}/{\sum_{s=1}^{s^{\rm max}}{s n_\infty (s,p)}}$, with $\tau$ and $\sigma$ defined in equations~(\ref{eq-tau}) and (\ref{eq-sigma}), 
should to match equation~(\ref{eq-gamma}).
Combining these results with equations~(\ref{Kinf}) and (\ref{eq-alpha}) delivers the scaling relations
\bea
\beta&=&\frac{\tau-2}{\sigma}, \label{11} \\
\gamma&=&\frac{3-\tau}{\sigma}, \label{12} \\
2-\alpha&=&2\beta+\gamma, \label{13} \\
2-\alpha&=&d\nu.
\label{eq-hypesc}
\eea
The last {of these relations} is known as hyperscaling 
and is the only one to involve the space dimensionality~$d$.

We collect here the MFT values of the critical exponents listed above:
\be
 \alpha = -1, 
 \quad
 \beta = 1,
 \quad
 \gamma=1,
 \quad
 \tau=\frac{5}{2},
 \quad
 \sigma=\frac{1}{2}
 \quad
 \nu = \frac{1}{2}.
 \label{Molly}
\ee
These are all correct above the upper critical dimension for percolation theory and obey the above scaling relations except  hyperscaling.
The failure of hyperscaling is evident in that the RHS of equation~(\ref{eq-hypesc}) is $d$-dependent while $\alpha$ and the other MFT exponents are independent of $d$ so long as  $d$ is above the upper critical dimension $d_{\rm c}=6$.
{The failure (and restoration) of hyperscaling is a central theme of this paper.}


The MFT values of $\gamma$ and $\nu$ listed above are the same as those in the Ising model.
The reason for this lies in the fact that both percolation and the Ising model are related to Landau theory, the free energy for which  reads
\be
 F[\phi] = \int{\rd^dx }
 \left({f_0 + \frac{r}{2}\phi^2
 +\frac{u_3}{3}\phi^3
 +\frac{u_4}{4}\phi^4
 + \dots 
 }\right).
\label{flowers}
\ee
In the Ising case, the ``up-down'' symmetry demands that only terms of even power in $\phi$ contribute. 
Thus, there can be no  $\phi^3$ term.
We cut the $\phi^n$ expansion at the smallest positive term that maintains this symmetry and at the same time maintains stability. In the symmetric Ising case that is the $n=4$ term.

Percolation theory is quite different in that there are no spins. 
Instead, we simply deem sites that are separated by a single lattice bond as being linked. 
Since there is no ``up-down'' symmetry,  the $\phi^3$ term is allowed and the expansion can be stopped there.
This theory is actually related to the $q$-state Potts model in the limit $q\rightarrow 1$.
Above $d_{\rm c}$, both the Ising model and percolation theory are controlled, in RG terms, by the so-called Gaussian fixed point.
That is obtained by maintaining only terms up to the quadratic 
in equation~(\ref{flowers}) --- thus it is the same for both theories.

Above the upper critical dimension, higher-order terms (quartic in the Ising case and cubic for percolation) are irrelevant in the RG sense. 
However, this irrelevancy turns out to be dangerous in some cases --- such as for the specific heat.
This affects the scaling behaviour and necessitates
modified treatment of associated scaling dimensions.
This is where the concept of dangerous irrelevant variables, introduced by Michael Fisher in the 1980's, comes in \cite{FiHa83}.
We discuss these at length in references~\cite{ourreview,SciP} and briefly here in section~\ref{sec_3}.


But first, to complete the analogy with ordinary magnetic phase transitions, we observe that the scaling theory of percolation can be built from a homogeneity assumption. 
We illustrate this theory below $d_{\rm c}$, where it is well understood.
For that purpose, let us rewrite equation~(\ref{eq-tau}) in the form
\be
n_\infty(s,p)=\xi^{-(d+D_{\rm f})}{\mathscr N}\big(s/\xi^{D_{\rm f}}\big),
\ee
having used equation~(\ref{sigma}) for $\sigma$ and equation~(\ref{tau}) for $\tau$.
The quantity $sn$ scales with {$\xi^{-d}$}, as a free energy does, and, as promised, can be identified as such analogously.
The free energy density for clusters of size $s$
therefore reads as
\be
K_\infty {(s,p)} \sim \xi^{-d}{\mathscr K}\big(s/\xi(p)^{D_{\rm f}}\big).\label{eq-homf1}
\ee
The average cluster size follows from
\be
 S_\infty (p) \sim\int \rd s\ \! sK {(s,p)} =\xi_\infty^{2D_{\rm f}-d}\int \rd\varsigma\ \! \varsigma{\mathscr K}(\varsigma).\label{eq-scalingChi}
\ee
Identifying this with $S_\infty (p)\sim|p-p_\infty|^{-\gamma}$ in  equation~(\ref{eq-gamma}), one has $\gamma=(2D_{\rm f}-d)\nu$. 
This relation also shows that the analog of the magnetic field
for percolation is $h\sim s^{-1}$.
The  homogeneity assumption~(\ref{eq-homf1})
 is usually written in terms of  an arbitrary rescaling factor $b$ instead of the correlation length, and therefore scales as
\be K(s,p)\sim b^{-d}{\mathscr F}\big(\kappa b^{D_{\rm f}},\epsilon b^{1/\nu}\big),\label{Caroline}\ee
with $\kappa=s^{-1}-(s^{\rm max})^{-1}$ and $\epsilon=|p-p_\infty|$ which measure the distance to the critical point.

The behaviour of the percolation probability and the average cluster size now follow from $P(p)\sim\partial K/\partial \kappa$ and
$ S(p)\sim\partial^2 K/\partial \kappa^2$ and deliver $\beta=(d-D_{\rm f})\nu$ and the same expression for $\gamma$ as above.
In terms of renormalization-group scaling dimensions, usually (below $d_{\rm c}$),  the correlation length exponent~$\nu$ is related to the  thermal exponent via $y_t=1/\nu$.
 The exponent
$D_{\rm f}$ is similarly the equivalent of the RG magnetic scaling dimension, $y_h=D_{\rm f}$.


\section{Finite-size scaling in percolation theory}\label{sec2}


For finite-sized systems, critical singularities are smoothened out. For example, the susceptibility which diverges at the critical point for infinite systems, is rounded,  bounded and shifted for finite ones. 
This is usually understood by a simple argument contained in the scaling form (\ref{eq-scalingChi}).
The susceptibility, at probability $p$, of a system in the thermodynamic limit is determined (at least to leading singularity) by the correlation length at the same probability;
$ S_\infty(p)\sim \xi_\infty(p)^{\gamma/\nu}$.
This latter quantity behaves { like} the system size $L$ in the neighborhood of the transition, so that
\be
 S_L\sim L^{\gamma/\nu}\label{eq-chiL}
\ee
there. This form of finite-size scaling (FSS), if it holds above $d_{\rm c}$ with MFT exponents, is often called Landau FSS.
Of course, this relation, if it holds, has an associated amplitude  which should depend on $p$ so that
$ S_L(p)\simeq \Gamma(p)L^{\gamma/\nu}$. 
The amplitude $\Gamma(p)$ follows from the homogeneity assumption
\be
 S_L(p)=L^{\gamma/\nu} {\mathscr S}\big(L^{1/\nu}|p-p_\infty|\big),\label{eq-chiLp}
\ee
where 
$  {\mathscr S}(y)$ is the universal scaling function associated with the average cluster size. 
Two applications of this relation are commonly used.

First, equation~(\ref{eq-chiL}) is a special case of equation~(\ref{eq-chiLp}) evaluated at the percolation threshold $p_\infty$ where the reduced variable $y=L^{1/\nu}\epsilon$ is zero, $\Gamma_0={\mathscr S}(0)$: 
$S_L(p_\infty)\simeq \Gamma_0L^{\gamma/\nu}$.
This is often called {\em critical-point} (FSS).
Second, the pseudocritical threshold $p_L$ is the value of $p$ for which the susceptibility reaches its size-dependent bound.
 The scaling function 
${\mathscr S}(y)$ reaches a maximum value, say $\Gamma_{y_0}={\mathscr S}(y_0)$. There, the relation (\ref{eq-chiL}) holds also, in the form 
$ S_L(p_L)\simeq \Gamma_{y_0}L^{\gamma/\nu}$, and the value of $p_L$ is shifted from $p_\infty$ by a size-dependent amount
\be
p_L=p_\infty\pm y_0L^{-1/\nu}.
\ee
We  also define  the rounding $\Delta p_{\frac 12}=|p_{+\frac 12}-p_{-\frac 12}|$, 
where $p_{\pm\frac 12}$ are the two values of $p$ for which the function ${\mathscr S}$ reaches half its maximum height ${\mathscr S}(y_0)$, i.e., such that 
${\mathscr S}(L^{1/\nu}|p_{\pm{\frac 12}}-p_\infty|)=\frac 12 {\mathscr S}(L^{1/\nu}|p_L-p_\infty|)$. 
The scaling function ${\mathscr S}(y)$ is non-singular. One can therefore expand the left-hand side of the previous equation in the vicinity of $y_0$ which leads eventually to
\be
\Delta p_{\frac 12}\sim L^{-1/\nu},
\ee 
where the omitted prefactor is a combination of  ${\mathscr S}(y_0)$ and its successive derivatives.
This is  {\em pseudocritical-} {\em point} {FSS}.

FSS based on equation~(\ref{eq-chiLp})  is   very widely used for systems --- both spin models and percolation --- below their upper critical dimensions.
However, it is not quite correct above $d_{\rm c}$, as we shall see.

Percolation, of course, carries another element: the fractal dimension of the maximal clusters.
From equations~(\ref{eq-nu}) and (\ref{eq-sigma}), their mass for finite size systems is $s^{\rm max}_\infty \sim \xi_\infty^{1/ \nu \sigma}$, which, using the above prescription becomes
\be
 s^{\rm max}_L \sim L^{D_{\rm f}}
\ee
for finite size.
With equations~(\ref{11}) and (\ref{12}) these give
\be
 D_{\rm f} = \frac{\beta+\gamma}{\nu}
 \label{D1}
 \ee
 and, if hyperscaling holds this means
 \be
 D_{\rm f} = d-\frac{\beta}{\nu}.
 \label{D2}
 \ee
However, we are interested in high dimensions and we have indicated that the picture described above does not apply there, or, more precisely, that it does not {\em fully} hold there.

Equation~(\ref{D1}) with mean-field exponents from equation~(\ref{Molly}) tells us that $D_{\rm f}=4 \equiv D_c$, i.e., the value of the magnetic RG eigenvalue at the upper critical dimension. Hyperscaling in equation~(\ref{D2}), on the other hand, tells us that $D_{\rm f}=d-2$.
Of course we know that hyperscaling in its old form~(\ref{eq-hypesc}) does not hold above the upper critical dimension. 
If  $D_{\rm f}$ is ``stuck'' at $D_c=4$ it leaves room for a  proliferation of the biggest clusters. 
This was the prevailing understanding in the statistical physics community since the seminal work of Antonio Coniglio in 1985, and it is, at least partially confirmed for systems with free boundary conditions.
However, the picture coming from the more mathematically oriented community was more aligned with the non-proliferation scenario --- a single cluster of fractal dimension $D_{\rm f}=2d/3=D^*$ spans the system.  Here we borrow the star notation of 
reference~\cite{BNPY}.
Amnon Aharony and Dietrich Stauffer furthermore provided an analysis of the shift of percolation threshold in high dimensional systems and found that 
\be p_\infty=p_L+AL^{-2}+BL^{-d/3},\ee 
with $A=0$ for systems with periodic boundary conditions~\cite{AhSt95}.
We have summarized these two scenarios in reference~\cite{OurJPA}  with a scaling Ansatz for the density of finite clusters $n(s,p)$. 
This is a central quantity, because, as we said,  $sn_L(s,p)$   in percolation theory scales like the free energy density in ferromagnets.
The two scenarios are as follows.

\begin{description}
\item[Scenario 1.] {\em Proliferation of maximal clusters:} 
In 1985, building on previous work by   Aharony and collaborators \cite{AhGe84},  Coniglio published a famous paper providing a geometric explanation for the breakdown of hyperscaling above the upper critical dimension for percolation \cite{Conig85}. 
His crucial insight was that spanning clusters (we call them maximal clusters)  can interpenetrate, allowing their density to diverge faster than would be expected otherwise. Quoting Coniglio:
\begin{quotation}
``Note that (\dots)  we expect at $p_\infty$ an infinite number of infinite clusters of zero density with fractal dimension 4.''
\end{quotation}
To allow for this,  
the free energy  obeys
\be
K(s,p,L)=b^{-(d-X)}{\mathscr F}\big(b^{D_c}\kappa ,b^{1/\nu}\epsilon,bL^{-1}\big).
\label{eq-scaling-n-Coniglio}
\ee
Here, $D_c$ is the fractal dimension of the maximal clusters as in equation~(\ref{Caroline}) 
and $X$ is a new exponent that accounts for interpenetration. It also measures the proliferation of spanning clusters, the number of which scales in $L^X$ in a finite system of linear size $L$.

Throughout the years, numerical results supportive of Coniglio's scenario for systems with free boundary conditions (FBC) were published, where {$D_c=4$} is fixed at its value at $d_{\rm c}$ as in equ\-ation~(\ref{D1}) and $X=d-6$ (see reference~\cite{OurJPA} for a review).
It should noted that the three exponents appearing in equation~(\ref{eq-scaling-n-Coniglio}) are precisely such that {\em standard} Landau FSS holds, i.e., FSS with MFT exponents.

A consequence of equation~(\ref{eq-scaling-n-Coniglio}) is that the scaling of the maximal cluster size $s^{\rm max}$ is
\be
s^{\rm max}(p)=L^{-X}\left(L^d\frac{\partial K}{\partial \kappa}\right)_{s=s^{\rm max}}=L^{D_c}\mathscr{ S}(L^{1/\nu}|p-p_\infty|).
\label{eq-smaxConiglio}\ee
The prefactor $L^{-X}$ accounts for the fact that  $s^{\rm max}$ refers to a single cluster (unique among the $L^X$ percolating clusters) and the ordinary volume term $L^d$ converts the density (per site) to an extensive quantity.

\item[Scenario 2.] {\em Non-proliferation of maximal clusters:}
An alternative, or parallel, theory emerged from the mathematical community.
This relates to Erd{\H{o}}s-R{\'{e}}nyi random graphs.
Already in the 1980's and early 1990's, B{\'{e}}la B
Bollob{\'{a}}s (1984) and Tomasz {\L}uczak (1990) proved that with $V$ sites in the graph, the giant component has $V^{2/3}$ sites at a certain probability \cite{Bo84,Lu90}. 
Moreover, the scaling window associated with this probability is of size $V^{-1/3}$. 
If percolation on finite graphs is like Erd{\H{o}}s-R{\'{e}}nyi random graphs, one expects $D_{\rm f}=D^*=2d/3$ and a rounding exponent of $d/3$.
In a series of papers in the 2000's, the mathematical community was able to show that this scenario indeed applies for high-dimensional percolation on a torus (i.e., PBCs).
For a review, we refer the reader to the lecture notes of
Markus Heydenreich and Remco van der Hofstad in reference~\cite{HH}.
Because of the similarities between the lattice and Erd{\H{o}}s-R{\'{e}}nyi set-up, the scaling is called random-graph asymptotics.

\end{description}

	 Nevertheless, there are other  options that one should explore. One of these (not hitherto considered) is non-proliferation of maximal clusters and Gaussian fixed point scaling. This will be explained later, but let us just mention the key assumption implied by this third scenario:

\begin{description}
\item[Scenario 3.] {\em Gaussian fixed point scaling:}
This possibility suggests, instead of equation~(\ref{eq-scaling-n-Coniglio}), 
\be
K(s,p,L)=b^{-d}{\mathscr F}(b^{d/2+1}\kappa,b^{2}\epsilon,bL^{-1}),
\ee
leading to  $s^{\rm max}(p_\infty)\sim L^{d/2+1}$.
\end{description}

Therefore, although the singularities described in equations~(\ref{eq-beta})--(\ref{eq-sigma}), and~(\ref{eq-alpha}) 
are quite valid in the thermodynamic limit, with the MFT values of the critical exponents (\ref{Molly}),  and although the above two scenarios were quite valid in their own rights, FSS was  puzzling.
There were two conflicting theories, each of which violated standard finite-size scaling and  hyperscaling as they emerge from the RG proper. 
With proliferation and non-proliferation depending on boundary conditions, universality itself was also under threat.
We shall see that with superlinear correlation length, the RG survives these challenges and emerges as a very robust descriptor of scaling in high dimensions as well as in low ones. There still remain open questions that we summarize at the end of this paper.

\section{Testing the theories numerically}
\label{sec_3}

We tested these scenarios against numerical simulations for two types of boundary conditions, PBCs and FBCs, and for $d=7$ and $d=8$ which both lie above the upper critical dimension in the MFT regime.

Scaling is essentially described by the homogeneity assumption for the density of finite clusters $K(s,p)$ which is characterized by three main quantities: the fractal dimension of maximal clusters, the thermal eigenvalue and the proliferation exponent.
The only one of these that is capable of distinguishing between all three scenarios is the fractal dimensionality of the largest cluster.
As a  preliminary study,  we therefore present here results for this quantity, $s_L^{\rm max}$.

In this timely report we do not aim to break the record for accuracy of critical exponents or the locations of the critical and pseudocritical points 
---  Peter Grassberger had that record for almost twenty years~\cite{Grassberger} and it was refined only  recently~\cite{Mertens}.
Instead, as discussed, our aim is to compare three different theoretical scenarios.
We determine the size of the maximal clusters for various values of $p$  using a standard finite-size fitting protocol. 
We do that using $L$ values ranging from $L=3$ to $L=30$ for $d=7$ and $L=3$ to $L=20$ for $d=8$. 

\begin{figure}[ht]
\begin{center}
\includegraphics[width=11cm]{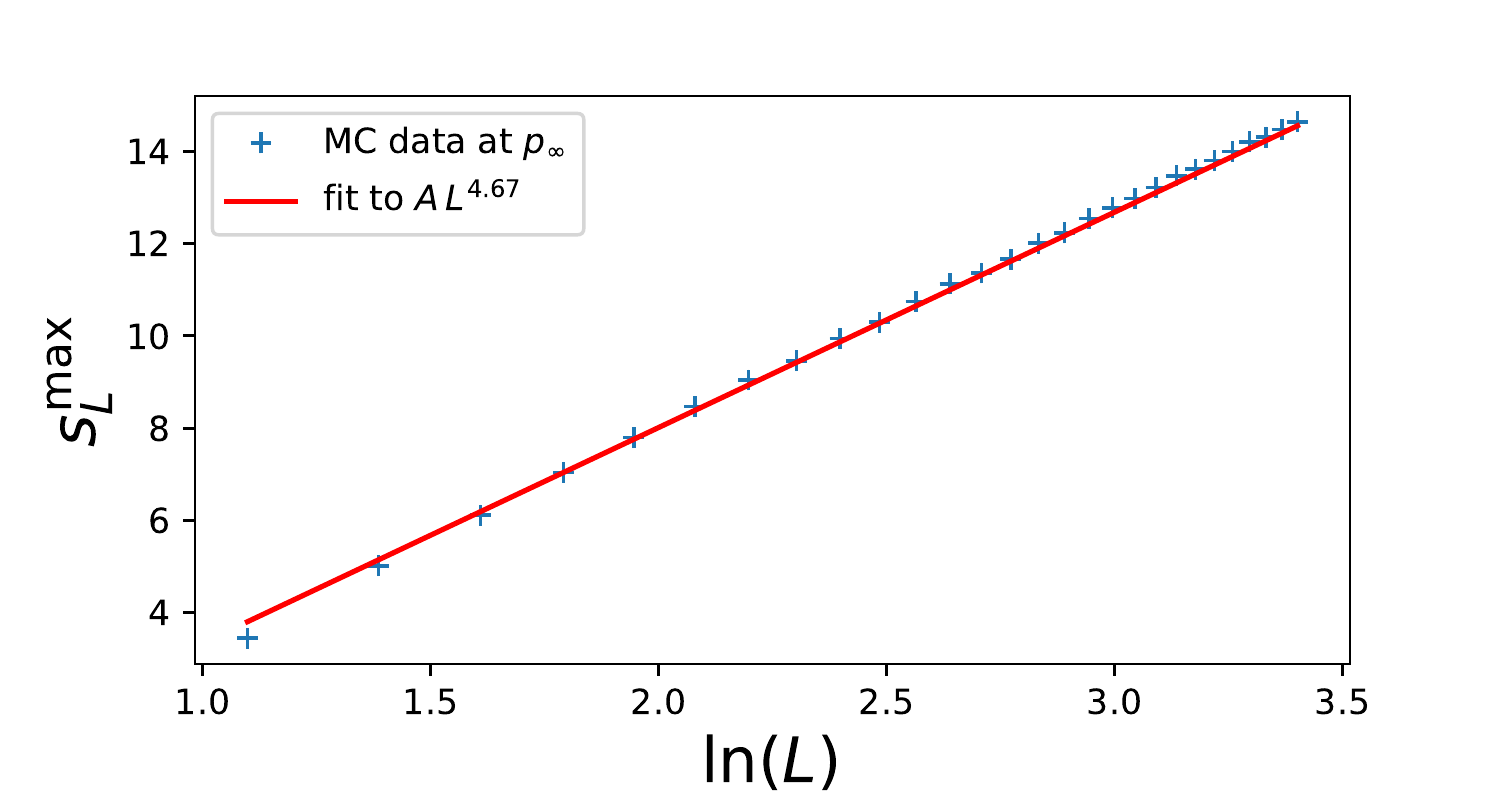}
\end{center}
\caption{(Colour online) FSS of data for $s^{\rm max}(p_\infty)$ plotted against $L$ 
(in the range $L=3$ to $L=30$)
for PBC in $d=7$. 
The MC  data (symbols) are fitted to $s^{\rm max}(p_\infty) =A \ \!L^{4.67}$.
}\label{figa}
\end{figure}

 The power law behaviour of the quantity $s^{\rm max}(p_\infty)$ evaluated at the  percolation threshold is then analysed. Results are shown as an example in figure~\ref{figa} 
 (for PBC at $d=7$, further cases are shown later in figure~\ref{fig1} and figure~\ref{fig2})
where
a log-log plot of the MC data for $s^{\rm max}_L(p_\infty)$ plotted against $L$ 
displays a leading power-law behavior compatible with 
$s^{\rm max}(p_\infty)\sim L^{D_{\rm f}}$ with $D_{\rm f}^{\rm FBC}=4$ and
$D_{\rm f}^{\rm PBC}=2d/3$ (only this second case is shown in figure~\ref{figa}).

\begin{figure}[ht]
\begin{center}
\includegraphics[width=11cm]{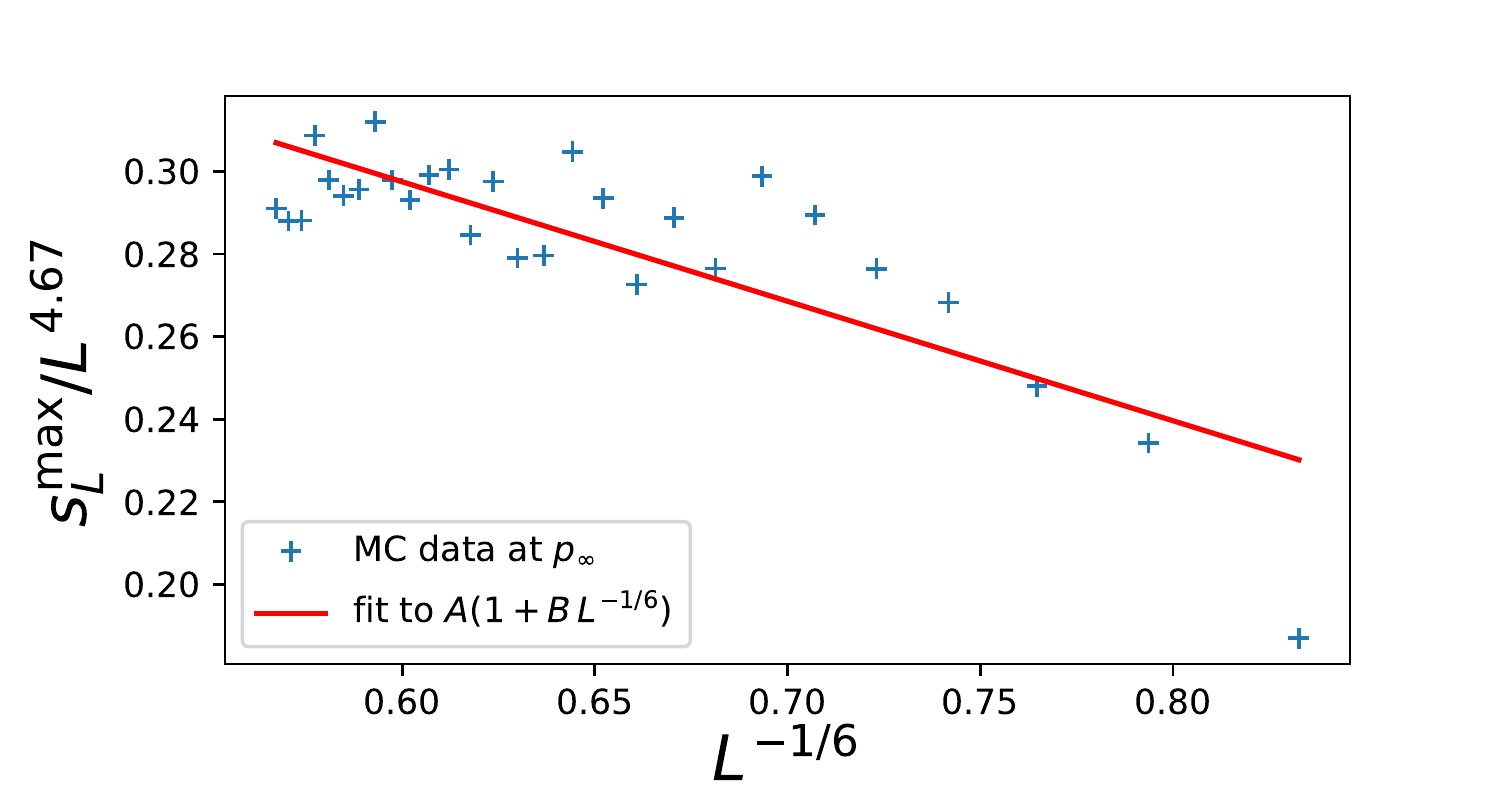}
\end{center}
\caption{(Colour online) FSS of data for $s^{\rm max}(p_\infty)$ rescaled by the leading power law $L^{4.67}$, plotted against $L^{-1/6}$ 
(in the range $L=3$ to $L=30$)
for PBC in $d=7$. 
The MC  data (symbols) are fitted to (solid red line) 
$s^{\rm max}(p_\infty) / L^{4.67}\simeq A(1+BL^{-1/6})$.
}\label{figb}
\end{figure}

\begin{figure}[h]
	\begin{center}
		\includegraphics[width=5.5cm]{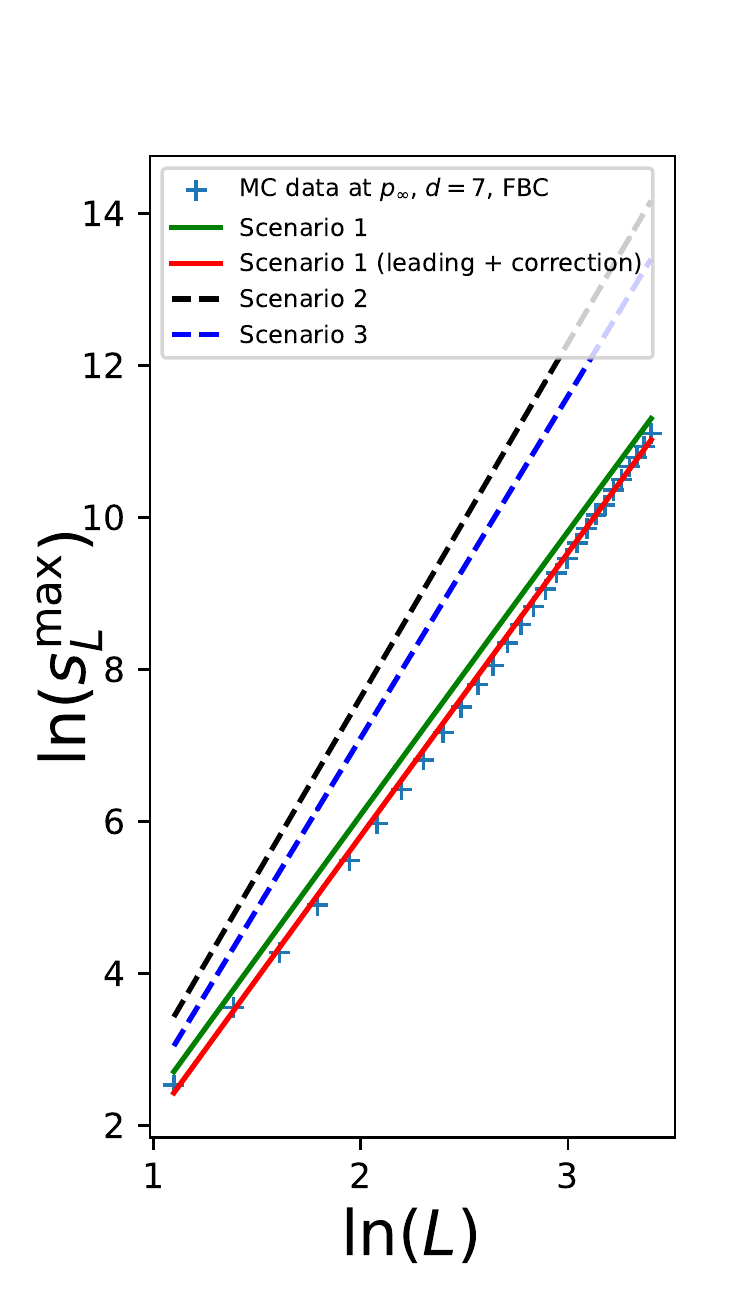}
		\includegraphics[width=5.5cm]{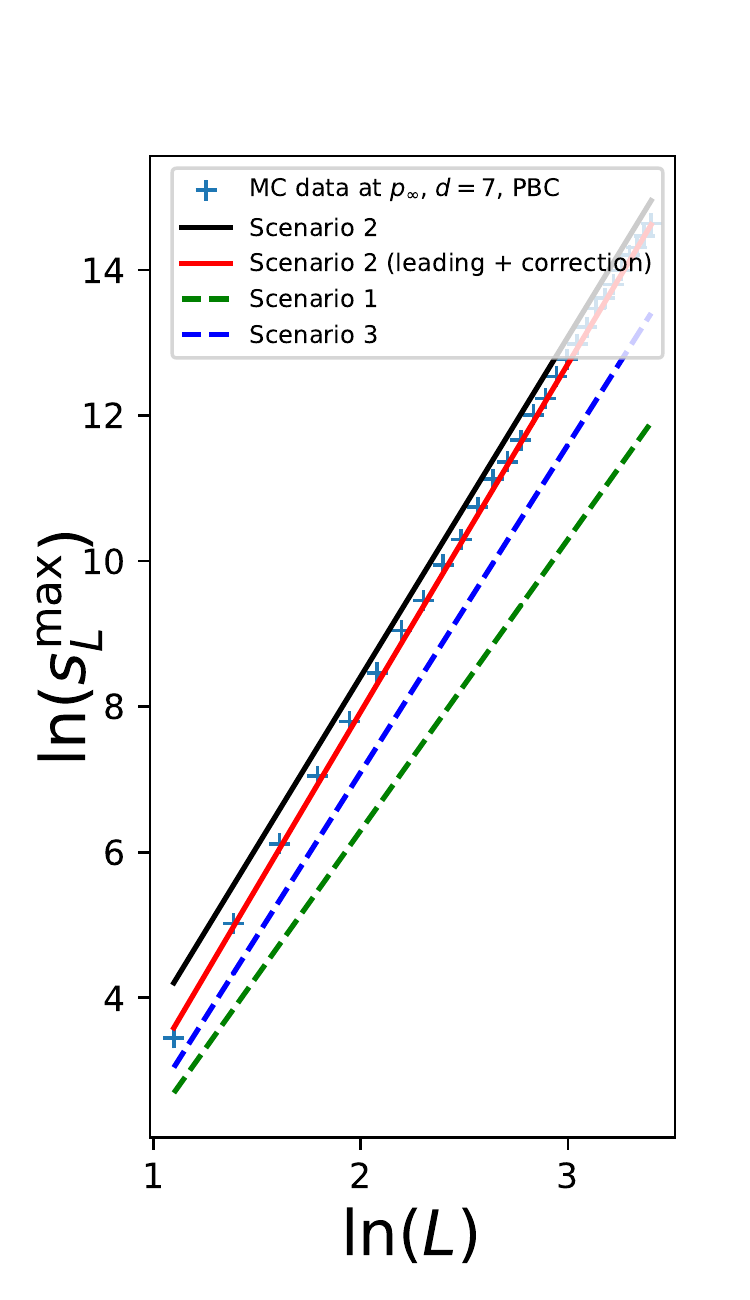}
	\end{center}
	\caption{(Colour online) FSS of data for $s^{\rm max}(p_\infty)$ plotted against $L$ 
		(in the range $L=3$ to $L=30$) for FBC (left-hand pannel) and
		for PBC (right-hand pannel) in $d=7$. 
		Symbols are the  MC  data. 
		The full and dashed lines (shifted along the vertical axis for clarity) are $\ln \big(L^{D_{\rm f}}\big)$ with 
		$D_{\rm f}=4$ (scenario 1), 
		$D_{\rm f}=4.67$ (scenario 2),  and
		$D_{\rm f}=4.5$ (scenario 3).
		For FBC, the leading behavior (solid green line) is given by scenario 1, while for PBC, this is scenario 2  (solid black line). Dashed lines  are the two scenarios which are not compatible with the MC data (the slopes are inconsistent). 
		The fit including the correction to scaling for each case (solid red line) is $\ln \big(s^{\rm max}_L \big)\simeq L^{D_{\rm f}}\big(1+AL^{1-d/6}\big)$ with the appropriate $D_{\rm f}=4$ and $4.67$ for FBC and PBC, respectively. }\label{fig1}
\end{figure}

\begin{figure}[h]
\begin{center}
\includegraphics[width=5.5cm]{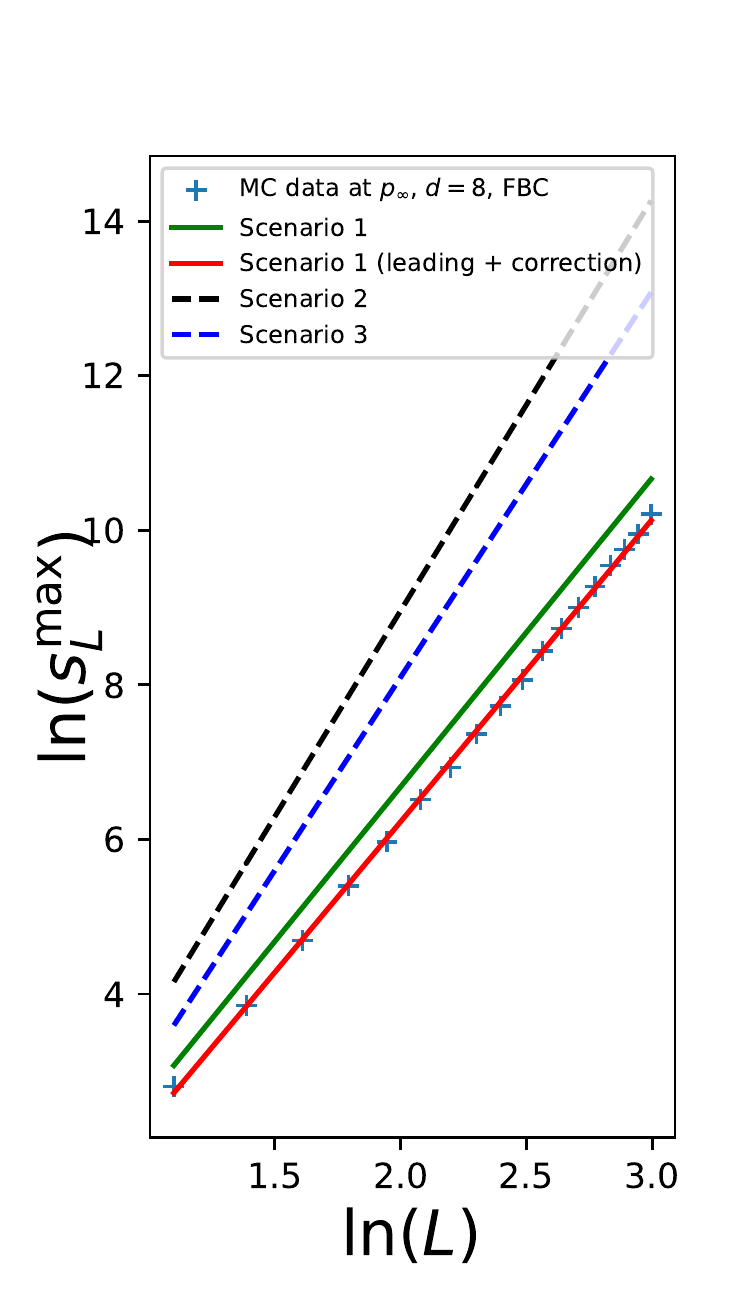}
\includegraphics[width=5.5cm]{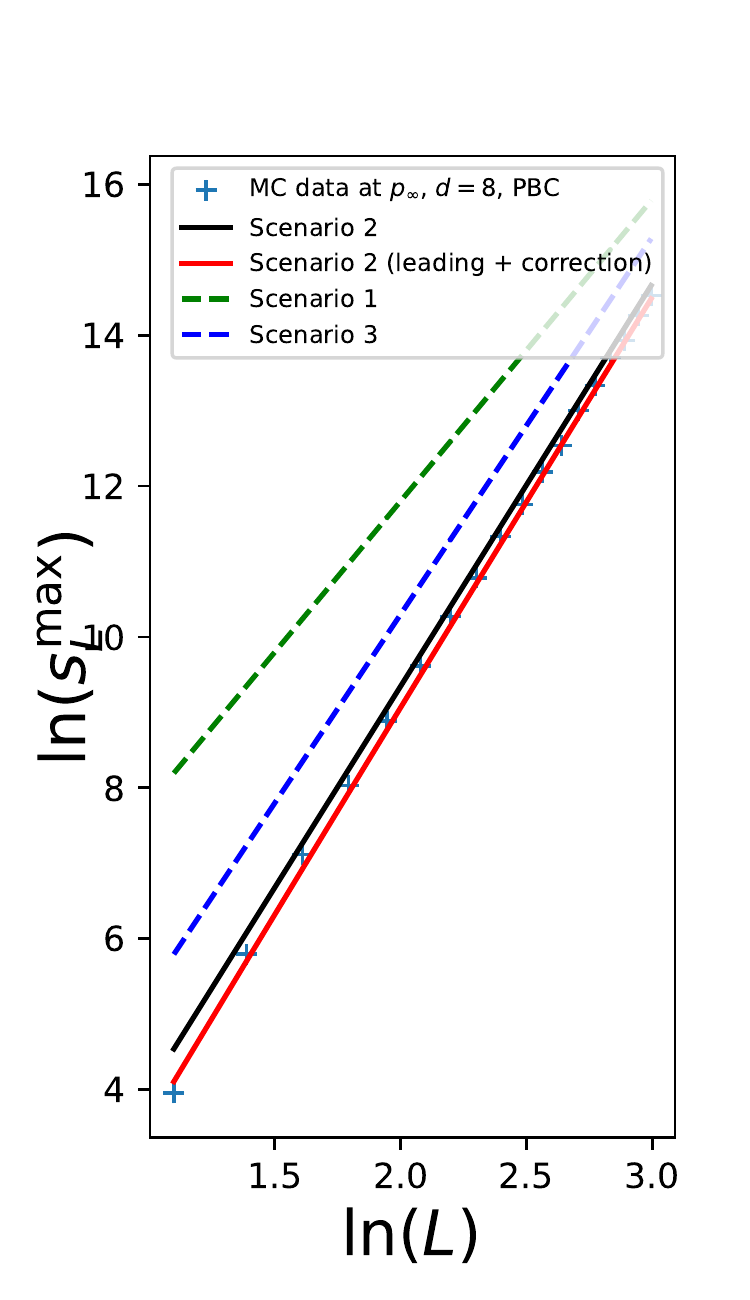}
\end{center}
\caption{(Colour online) Same as figure~ \ref{fig1} for $d=8$ 
(in the range $L=3$ to $L=20$).
The values of $D_{\rm f}$ are $4$ and $5.33$ for FBC and PBC, respectively.
}\label{fig2}
\end{figure}

The corrections to scaling are discussed later in the following section, but we anticipate here the analysis in  figure~\ref{figb}. There, still in the case $d=7$, PBC,  we show that the MC data are compatible with 
$s^{\rm max}_L(p_\infty)\simeq A L^{D_{\rm f}}(1+BL^{1-d/6})$, i.e., a correction to scaling in $L^{1-d/6}$ besides the leading power law.

The three scenarios are then compared and the correction to scaling is also incorporated in the best scenario for each choice of boundary conditions.
For that purpose, we plot the rescaled MC data  $\ln \left[s_L^{\rm max}(p_\infty)/L^{D_{\rm f}}\right]$ against $\ln L$ and draw the different scenarios for comparison. This is done for $d=7$ with FBC in figure~\ref{fig1} (upper panel) and with PBC in figure~\ref{fig1} (lower panel). The case $d=8$ is displayed in figure~\ref{fig2}, with FBC in the upper panel and with PBC in the lower panel.
It appears clearly that the FBC results are compatible with scenario 1 (proliferation of maximal clusters, $D_{\rm f}=4=D_c$) and that  PBC results, as announced in figure~\ref{figa}, agree with scenario 2 (non-proliferation of maximal clusters, $D_{\rm f}=2d/3=D^*$). The third scenario (Gaussian fixed point scaling) does not seem to be confirmed in any of the two boundary conditions. 

With this material at hand, i.e., the scaling of $s^{\rm max}(p_\infty)$, we can make further deductions. The homogeneity assumption of equation~(\ref{eq-smaxConiglio})  can be generalized to the form
\be
s^{\rm max}_L(p)=L^{D_{\rm f}}
\mathscr{S}(L^{y}|p-p_\infty|),
\label{eq-smaxGen}
\ee 
with $D_{\rm f}$ and $y$ free exponents in order to fit any of the three scenarios.
This expression demands that $\mathscr{S}(x)\sim x^{- 1/\sigma}$ with $\sigma=\frac 12$ to recover the MFT exponent in the thermodynamic limit $L\to \infty$. This requires the further condition
\be
y=\sigma D_{\rm f}
\ee 
to cancel the $L$-dependence.
FBC results are compatible with the proliferation of maximal clusters (scenario 1) for which one thus has
\be y_c=\sigma D_c=2=\frac 1\nu, \ee
in agreement with Coniglio's picture
while PBC results have shown that $D_{\rm f}=D^*=2d/3$ and they lead to
\be
y^*=\sigma D^*=\frac d3,\label{eq-ystar}
\ee 
in agreement with random-graph asymptotics.

%

\section{Q-adjusted finite-size scaling in percolation theory}
\label{sec_4}

The attentive reader will notice that the situation is reminiscent of the case of the $\phi^4$-model. There, the MFT exponents $\alpha=0$, $\beta=1/2$, $\gamma=1$, $\delta=3$, $\eta=0$ and $\nu=1/2$, describe the singularities of the physical quantities above the upper critical dimension, $d_{\rm c}=4$, in the thermodynamic limit. This is at first glance surprising, since the stable RG fixed point there is the Gaussian one (GFP) treating only relevant terms  and should imply exponents $\alpha$, $\beta$ and $\delta$ which differ from the MFT ones. 
As stated earlier, the solution of the puzzle was given by Michael Fisher in terms of the danger of the irrelevant variable $u$ (the coefficient of highest degree in the Landau expansion of the free energy). 
This variable contaminates the homogeneity assumptions in a form which is compactly written in terms of a new exponent, $\koppa=d/4$, and the full scenario is compatible with
\bea
&&f(h,t)=b^{-d}{\mathscr F}\big(b^{d/2+\smallkoppa}|h|,b^{2\smallkoppa}|t|,bL^{-1}\big),\label{eq-f-coppa}\\
&&\xi(h,t)=b^{\smallkoppa}{\mathscr X}\big(b^{d/2+\smallkoppa}|h|,b^{2\smallkoppa}|t|,bL^{-1}\big).\label{eq-xi-coppa}
\eea
The first of these equations is essentially the original form of FSS taking Fisher's dangerous irrelevant variables into account~\cite{FiHa83}.
The second is a recent extension that allows for superlinearity of the correlation length~\cite{koppa}.
We find it expedient to refer to this fuller scaling picture as Q in order to distinguish it from previous theories in which $\koppa$ was effectively set to 1. 
It also serves to distinguish from Gaussian which we refer to as G and which remains part of the full scaling picture.

Indeed, a careful analysis of order parameter Fourier modes (standing waves for PBCs  and sinusoidal waves for FBCs) shows that those modes which are contaminated by the dangerous irrelevant variable are those which can develop non-zero expectation values in compatibility with the boundary conditions.
We refer to these as Q-modes in~\cite{ourPRL}. 
The other modes are entirely controlled by the GFP and described by 
homogeneity assumptions associated with Gaussian RG eigenvalues $y_t$ and $y_h$:
\bea
&&f^{\rm G}(h,t)=b^{-d}{\mathscr F}^{\rm G}\big(b^{d/2+1}|h|,b^{2}|t|,bL^{-1}\big),\label{eq-f-coppaG}\\
&&\xi^{\rm G}(h,t)=b{\mathscr X}^{\rm G}\big(b^{d/2+1}|h|,b^{2}|t|,bL^{-1}\big).\label{eq-xi-coppaG}
\eea
Here, the superscript G is for Gaussian and we refer to G-modes in  Fourier terms.\footnote{In the renormalization-group set-up, the exponents find their origins in scaling dimensions. 
The arguments of the Gaussian functions $f^{\rm{G}}$ and $\xi^{\rm{G}}$ are $y_h=d/2+1$ and $y_t=2$ for both the Ising model and for percolation.
These neglect the danger  pointed out by Fisher of the irrelevant quartic and cubic terms in equation~(\ref{flowers}).
After accounting for these, they are replaced by $y_h^*$ and $y_t^*$ which differ for each model (because they come from quartic or cubic terms, respectively).
In terms of $\koppa$, they are $y_h^*(d)=y_h(d)+(\koppa-1)=\koppa y_h(d_{\rm c})$ and $y_t^*(d)=y_t(d)+2(\koppa-1)=\koppa y_t(d_{\rm c})$.
Thus, setting $\koppa$ is a measure of the danger of Fisher's dangerous irrelevant variables and setting $\koppa =1$ converts the corrected scaling forms  (\ref{eq-f-coppa}) and  (\ref{eq-xi-coppa}) to 
(\ref{eq-f-coppaG}) and  (\ref{eq-xi-coppaG}), respectively.}

Extensive numerical simulations of high-dimensional Ising models, and of Ising models with long-range interactions above their upper critical dimensions~\cite{ourEPJB,ourPRL}, have been performed to check the FSS predictions of equations~(\ref{eq-f-coppa}) and (\ref{eq-xi-coppa}) (see reference~\cite{ourreview,SciP} for a review and reference~\cite{Lundow2021} for the latest impressive simulations).
Our picture now is that Q scaling holds at the pseudocritical points for both PBC's and FBC's in various spin models. 
At the infinite-volume critical temperature on the other hand, only FSS in systems with PBC's obey equations~(\ref{eq-f-coppa}) and (\ref{eq-xi-coppa}). 
Systems with FBCs escape this scenario due to an anomalous shift of the pseudocritical temperature 
$|T_L-T_\infty|  \sim L^{-\lambda}$ with $\lambda =2$, instead of 
$\lambda=2\koppa$ for PBCs. 
The rounding in both cases obeys $\Delta T_{\frac 12}\sim L^{-2\smallkoppa}$.
Therefore, the infinite-volume critical point for FBCs
is far outside the pseudocritical scaling window and the finite system  at $T_\infty$ is effectively deep
in the disordered phase. 
The dangerous variable cannot contaminate the order parameter, and a scaling consistent with equations~(\ref{eq-f-coppa}) and (\ref{eq-xi-coppa}) there may be more likely, although the situation drawn from the numerical results is
thus far inconclusive~\cite{ourreview,SciP,ourPRL,Brankov,BNPY,LuMa11,LUNDOW2014249,WiYo14,Lundow2021}.

While the above numerical tests pertain to the Ising model or variants of it, besides the preliminary results reported above, no  tests of the Q-framework have been carried out for percolation.
We note that $D^*=2d/3$ coincides with $\koppa D_c$ and $y^*=d/3$ with $\koppa y_c$, i.e., they are what the Q picture predicts. This comes in full support of equation~(\ref{eq-f-coppa}) for percolation [with $K(s,p)$ instead of $f(h,t)$], but the case of equation~(\ref{eq-xi-coppa}), which is essential to establish the superlinearity of the correlation length, is still to consolidate.
Let us assume a scaling of the correlation length in the form similar to equation~(\ref{eq-smaxGen}),
\be
\xi_L(p)=L^a{\mathscr X}(l^y|p-p_\infty|),
\ee 
with, again, $a$ free, but $y$ fixed by the previous constraints, either to $y_c$ in scenario 1, or to $y^*$ in scenario 2. Since $\xi_\infty(p)\sim |p-p_\infty|^{-\nu}$ in the thermodynamic limit, we demand ${\mathscr X}(x)\sim x^{-\nu}$ and the $L$-dependence disappears when $a-y\nu=0$.
For FBCs it follows that 
$a_c=1$ while for PBCs it confirms Q again, with 
$a^*=d/6=\koppa$ (instead of $d/4$ for the Ising model).
Equation~(\ref{eq-xi-coppa}) is now reinforced for PBCs in the case of percolation in the form
\be
K(s,p)=b^{-d}{\mathscr F}(b^{D^*}\kappa,b^{y^*}\epsilon).\label{eqKSc2}
\ee

We can now explore the question of the hyperscaling relation w.r.t. these two scenarios.
Equation~(\ref{eq-scaling-n-Coniglio}) for the free energy density in Coniglio's approach, with the  choice $b=\epsilon^{-1/y_c}$ for the rescaling factor, leads to
\be
K(p)\sim\epsilon^{({d-X})/{y_c}}\sim\epsilon^{2-\alpha},
\ee
hence,
\be
\alpha=2-\frac{d-X}{y_c}.
\ee
Q scaling on the other hand (scenario 2) with 
$b=\epsilon^{-1/y^*}$ in equation~(\ref{eqKSc2}) leads to
\be
K(p)\sim\epsilon^{{d}/{y^*}}\sim\epsilon^{2-\alpha},
\ee
which then implies 
\be
\alpha=2-\frac{d}{y^*}.
\ee
To summarise, either traditional hyperscaling is broken by proliferation, with $X=d-(2-\alpha)/\nu = d-6$ (FBC case), or it is broken by superlinearity of the correlation length with $\koppa=d/d_{\rm c}=d/6$ (PBC case).
However, there appears  to be no room for both scenarios simultaneously so that one does not have proliferating clusters at the pseudocritical point or at the critical point with PBCs.

In section~\ref{sec2}, we have introduced a third scenario according to which the GFP could control the behaviour, without dangerous irrelevant variables.
The preliminary numerical results that we have produced seem to rule out this option as well.
This is  surprising, because this seems to be conflicting with what is observed in spin models.
Per Lundow and Klas Markstr\"om  have studied  this problem
expertly and intensively --- with corrections ---  in the case of the IM in $d=5$ with FBCs, reaching far large systems (up to $L=160$ in \cite{LUNDOW2014249}) and they obtained very accurately leading FSS behaviours and corrections to scaling:
\bea
&&\chi_L(T_c)=0.817\ \!L^2+0.083\ \!L,\label{E177}\\
&&m_L(T_c)=0.230\ \!L^{-3/2}+1.101\ \!L^{-5/2}-1.63\ \!L^{-7/2}.\label{E187}
\eea
The leading singularity of the magnetization is clearly compatible with the GFP value $2-d/2$ rather than with Landau FSS which would predict an exponent $-\beta/\nu=-1$.
This calls for further simulations, investigating the FSS behaviours of other physical quantities.


\begin{table}[h]
\caption{Scaling predictions for percolation, expressed in terms of FSS as $\xi_L\sim L^{\smallkoppa}$, $s^{\rm max}_L \sim L^{D}$, $N_L \sim L^X$, 
$S_L\sim L^{2D_{\rm f}-(d-X)}$, and $P_L\sim L^{-(d-X-D_{\rm f})}$. }

\vspace{0.25cm}
\begin{center}
\begin{tabular}{l|ccccc} 
 \hline \hline   
Scenario              
&
$\phantom{{\displaystyle\sum}}$ 
  $\koppa$                              
& $D$
& $X$
& $2D_{\rm f}-(d-X) $ 
& $d-X-D_{\rm f}$ 
\\
 \hline   
1. 
Proliferation
& 
$\phantom{{\displaystyle\sum}}$  
  1 
& $4$ 
& $d-6$ 
& 2&2 
\\ 
 \hline
2. 
Superlinearity
& $\phantom{{\displaystyle\sum}}$ 
$\displaystyle {d}/{6}$ 
& $\displaystyle {2d}/{3}$ 
& ${\displaystyle{0}}$  
& $\displaystyle d/3$ 
& $\displaystyle d/3$ 
\\  \hline
3. Neither
& 
$\phantom{{\displaystyle\sum}}$  
1 
& ${\displaystyle  (d/2)+1}$ 
& $0$ 
& $2$ 
& $\displaystyle (d/2)-1$ 
\\
\hline
	  \hline
\end{tabular}
\end{center}
\label{table1}
\end{table}

In table~\ref{table1}, we present predictions for several relevant quantities from the three scenarios we have discussed above.
The first is Coniglio's  scenario which we presently believe to be valid at the percolation point for FBCs.
This explains the failure of hyperscaling in its old form as due to proliferation of infinite clusters whose fractal dimensionality is ``stuck'' at $D=4$, irrespective of lattice dimensionality, but produces purely Landau FSS.
The second comes from  Q-scaling and is  expected at critical point for PBCs. 
In this case, hyperscaling is repaired by the introduction of $\koppa$.
The third scenario is what would come from the old hyperscaling relation (\ref{eq-hypesc}). In this case, $X=0$ and $\koppa=1$ so that neither come to the rescue.
Another open question is the behaviour at the pseudocritical threshold. In spin models, both PBCs {\em and} FBCs are controlled by Q scaling. What happens in percolation is still an open question.

As anticipated in figure~\ref{figb}, the best scenario in each case can still be improved if one takes into account corrections to scaling. For that purpose, the RG analysis proposed by Luijten \cite{Luijten,SciP} for the Ising model above $d_{\rm c}=4$ with PBC can be extended to the case of percolation ($d_{\rm c}=6$), leading here to
\bea 
s^{\rm max}(p,u)&=&L^{\frac d2+\smallkoppa}{\mathscr S}\left(L^{2\smallkoppa}(p-p_\infty) u^{(d_{\rm c}-2)/d_{\rm c}}-{\rm const} \ L^{[{ 4(d_{\rm c}-d) }/ {d_{\rm c}(d_{\rm c}-2)}]}\right)\nnb\\
&\simeq&L^{\frac d2+\smallkoppa}(1+A L^{1-\smallkoppa}+ \dots).
\eea
Here, $\koppa=d/d_{\rm c}$ takes the value $d/6$.
The corresponding fits presented in figures \ref{fig1} and \ref{fig2} (lower curves) were discussed earlier. Though the form of the correction relies on the RG analysis which is otherwise confirmed in PBC, it seems to fit also the numerical data in FBC, but we have no strong argument in support there.

\section{Conclusions }
\label{sec_5}

Here we have followed a tradition of importing insights developed in the context of ferromagnetism to further our understanding of percolation theory in high dimensions.
In principle, both should be covered by the renormalization group in the thermodynamic limit and for finite size, for periodic and free boundaries. 
However, both have resisted the attempts to be fully brought into the RG fold for a very long time. 
Finite size scaling, hyperscaling and universality do not hold in their traditional forms. 
Ingenious and creative solutions  were applied --- Fisher's dangerous irrelevant variables, Binder's thermodynamic length, Coniglio's proliferation of infinite clusters and so on.
However, all of these solutions were created under the assumption that the correlation length cannot exceed the length. 
This assumption seemed sound given  the coincidence of Gaussian scaling in the correlation sector with mean field --- why fix something that does not appear to be broken?
Thus, ad hoc solutions appeared to have closed the question.
Yet a discomfort persisted - why would different boundary conditions deliver different results? Why would universality not reside in the critical point for finite size when it did in the thermodynamic limit?
Thus, previous mathematical results were sometimes neglected or forgotten by the numerically-focused community, e.g., {\'{E}}douard Br\'{e}zin's observation that superlinearity of the correlation length in ferromagnets \cite{Br82} and rigorous proofs of random graph  asymptotics in percolation \cite{HH}.
These mathematical results pertain mostly to PBCs. 

Superlinearity of the correlation length finally allowed for the renormalization group in reference~\cite{ourNPB} and related papers, while maintaining the Gaussian scaling in the background \cite{ourPRL}. 
These considerations were finally imported to percolation in reference~\cite{OurJPA}.
Here we have shown that this opens up new possibilities for what is a rich scaling landscape in high dimensions.
We take a first attempt at testing some of these possibilities through numerical simulations.
We find that both the proliferation and random-graph  scenarios are valid --- each in their different regimes.
But most importantly, each is brought within the fold of renormalization group.


\section*{Acknowledgements}


Our friend (and former PhD mentor of one of us, BB) Lo\"\i c Turban, passed away in November 2022. Lo\"\i c was a generous and ethical scientist.
It was at his side that BB was introduced to the basics of percolation theory while still an undergraduate student.

We also dedicate this paper to freedom on which academic tradition stands.
We thank Julian Honchar and Yurij Holovatch, our Ukrainian collaborators and friends for numerous discussions. 
We would like to use the opportunity to express our unwavering support and friendship to them and to Maxym Dudka, Taras Holovatch, Olesya Mryglod, Marjana Krasnystka, Ihor Mryglod, Ulana Holovatch and so many others, and to the Ukrainian people.
Finally, TE and RK would like to thank the Squirrel, Rugby, for cheerfully accommodating the many discussions that formed the genesis of this paper.

This is a pleasure to submit this paper for our colleague Taras Bryk’s 60th birthday.

This work was supported by the Doctoral College for the Statistical Physics of Complex Systems, Leipzig-Lorraine-Lviv-Coventry, Franco-German University.


\newpage

\ukrainianpart
\title{50 років пошуку універсальності в перколяційній теорії вищих розмірностей}
\author{	Т. Елліс\refaddr{label1}, 
	Р. Кенна\refaddr{label2},
	Б. Берш\refaddr{label3}}
\addresses{
	\addr{label1} Дослідницький центр прикладної математики, Університет Ковентрі, Англія
	\addr{label2} Дослідницький центр прикладної математики, Університет Ковентрі, Англія та Проект L4, Ляйпціг-Лотарінгія-Львів-Ковентрі, Європа
	\addr{label3} Лабораторія теоретичної фізики та хімії,  CNRS - Університет Лотарінгї, 7019,	Нансі, Франція та Проект L4, Ляйпціг-Лотарінгія-Львів-Ковентрі, Європа
}
\makeukrtitle
\begin{abstract}
	Хоча в термодинамічній границі скейлніг добре описується в рамках теорії середнього поля, 
	для скінчених систем у вищих розмірностях він залишався незрозумілим. Це викликало певні питання щодо ефектив\-нос\-ті ренормгрупового підходу та таких базових понять, як універсальність, чисельний
	скінченорозмірний скейлінг та гіперскейлінг, які до останнього часу вважалися незастосовними вище верхньої кри\-тич\-ної вимірності.
	Значний теоретичний прогрес досягнуто у розв'язанні цих питань, що вже перевірено при чисельних моделюваннях спінових систем. 
	Цей прогрес грунтується на ідеї про суперлінійність кореляційної довжини --- понятті, яке довший час зустрічало спротив, але тепер є широко прийнятним серед наукової спільноти. У теорії перколяції виникають додаткові ускладнення, такі як розповсюдження взаємнопроникних кластерів, що, очевидно, протирічить висновкам, які випливають з асимптотики випадкових графів та є наслідком недостачі надійної статистики симуляцій.
	Обговорюються нещодавні теоретичні досягнення в теорії перколяції в рамках ренормгрупового підходу для вищих розмірностей,
	яка пристосовує суперлінійну кореляцію та робить більшість із вищезгаданих концепцій взаємно суміс\-ними за різних граничних умов. 
	Представлено результати числових розрахунків для вільних та періодич\-них граничних умов, які показують різницю між раніше конкуруючими теоріями. Незважаючи на певну фрагментарність, ці результати розрахунків методом Монте-Карло свідчать на користь
	нового підходу, на якому базується ренормалізаційна група і основні концепції, пов'язані з нею.
	
	\keywords перколяція, фазовий перехід, критичні показники, верхні критичні розмірності, співвідношення гіперскейлінгу, маржинальна нерелевантна змінна
\end{abstract}

\lastpage

\begin{thebibliography}{99}

\bibitem{BH} 
 Broadbent S. R.,  Hammersley J. M.,
Math. Proc. Cambridge Philos. Soc.,
1957, {\bf{53}},  629--641,\\
\doi{10.1017/S0305004100032680}.
\bibitem{Stauffer}
Stauffer D.,
{Introduction to Percolation Theory}, 
Taylor and Francis, London, 1985,
\bibitem{EssG}
Essam J., Gwilym K. M.,
J. Phys. C: Solid State Phys., 1971, {\bf{4}},  L228,
\doi{10.1088/0022-3719/4/10/015}.




\bibitem{FK69}
 Kasteleyn P. W.,   Fortuin C. M.,
 Physica, 1974, {\bf 57},  536,
 \doi{10.1016/0031-8914(72)90045-6}.

%
%
%
%
%

\bibitem{Essam}
 Essam~J.~W.,
Rep. Prog. Phys., 1980, {\bf{43}},  833--912, \doi{10.1088/0034-4885/43/7/001}.

\bibitem{SciP}
Berche B., Ellis T., Holovatch Yu., Kenna R., 
SciPost Phys. Lect. Notes, 2022, {\bf 60},  1--44,\\ \doi{10.21468/SciPostPhysLectNotes.60}.



\bibitem{FiHa83}
 Fisher M. E., 
 In: Lecture Notes in Physics. Critical phenomena, Vol.~186,
 Hahne F. J. W. (Ed.), 
 Springer Verlag, Berlin, 1983,  1--139.


\bibitem{ourreview}
Kenna R., Berche B.,  
In: Order, Disorder, and Criticality: Advanced Problems of Phase Transition Theory, Vol.~4,
 Holovatch Yu. (Ed.),  World Scientific, Singapore, 2015, 1--54.


\bibitem{BNPY}
Binder K., Nauenberg M., Privman V.,  Young A. P., 
Phys. Rev. B, 1985, {\bf{31}},  1498--1502,\\ \doi{10.1103/PhysRevB.31.1498}.


\bibitem{AhSt95}
Aharony A., Stauffer D., 
Physica A, 1995, {\bf{215}},  242--246, \doi{10.1016/0378-4371(95)00034-5}.

\bibitem{OurJPA}
Kenna R., Berche B.,
J. Phys. A: Math. Theor., 2017,  {\bf{50}},   235001, \doi{10.1088/1751-8121/aa6bd5}.

\bibitem{AhGe84}
Aharony A., Gefen Y., Kapitulnik A.,
J. Phys. A: Math. Theor., 1984, {\bf{17}}, L197, \doi{10.1088/0305-4470/17/4/008}.


\bibitem{Conig85}
Coniglio A., 
In: Physics of Finely Divided Matter. Springer Proceedings in Physics, Vol. 5,
Daoud M., Boccara~N.~(Eds.),
Springer Verlag, Berlin, 1985, 84--101.



\bibitem{Bo84}
 Bollob{\'{a}}s B.,
Trans. Am. Math. Soc., 1984, {\bf{286}}, 257--274, \doi{10.1090/S0002-9947-1984-0756039-5}.

\bibitem{Lu90}
 {\L}uczak  T., Random Struct. Algorithms, 1990, {\bf{1}}, 287--310, \doi{10.1002/rsa.3240010305}.



\bibitem{HH}
 Heydenreich M., van der Hofstad R.,
Progress in High-Dimensional Percolation and Random Graphs,
Springer~Cham, Switzerland, 2017.

\bibitem{Grassberger}
 Grassberger P., 
Phys. Rev. E, 2003, {\bf{67}},  036101, \doi{10.1103/PhysRevE.67.036101}.

\bibitem{Mertens}
 Mertens S.,  Moore C., Phys. Rev. E, 2018, {\bf 98}, 022120, \doi{10.1103/PhysRevE.98.022120}.

\bibitem{Luijten}
Luijten E., Ph.D. Thesis,  Delft University, Netherlands, 1997.
\bibitem{koppa}
 Kenna R., Berche B., 
Condens. Matter Phys., 2013, {\bf 16}, 23601, \doi{10.5488/CMP.16.23601}.  

\bibitem{ourPRL}
Flores-Sola E. J., Berche B., Kenna R., Weigel M., 
Phys. Rev. Lett., 2016, {\bf{116}}, 11570,\\ \doi{10.1103/PhysRevLett.116.115701}.

\bibitem{ourEPJB}
Flores-Sola E. J., Berche B., Kenna R., Weigel M., 
Eur. Phys. J. B, 2015, {\bf{88}}, 28, \doi{10.1140/epjb/e2014-50683-1}.




\bibitem{Lundow2021}
Lundow P. H., 
Nucl. Phys. B, 2021, {\bf{967}}, 
115422, \doi{10.1016/j.nuclphysb.2021.115422}.




\bibitem{Brankov}
Brankov J. G.,  Danchev D. M.,   Tonchev N. S.,
{{Theory of Critical Phenomena in Finite-Size Systems: Scaling and Quantum Effects}}, 
World Scientific, Singapore, 2000.

\bibitem{LuMa11}
 Lundow P. H., Markstr{\"{o}}m K., 
Nucl. Phys. B, 2011, {\bf 845}, 120, \doi{10.1016/j.nuclphysb.2010.12.002}.


\bibitem{WiYo14}
Wittmann M., Young A. P., 
Phys. Rev. E, 2014, {\bf{90}}, 062137, \doi{10.1103/PhysRevE.90.062137}.


\bibitem{LUNDOW2014249}
 Lundow P. H., Markstr{\"{o}}m K., 
Nucl. Phys. B, 2014,  {\bf 889}, 249, \doi{10.1016/j.nuclphysb.2014.10.011}.



\bibitem{Br82}
 Br\'ezin E., 
 J. Phys., 1982,  {\bf{43}}, 15--22, \doi{10.1051/jphys:0198200430101500}.

\bibitem{ourNPB}
 Berche B., Kenna R., Walter J. C.,
 Nucl. Phys. B, 2012,  {\bf{865}}, 115--132, \doi{10.1016/j.nuclphysb.2012.07.021}.

\end{thebibliography}
\end{document}